\documentclass[10pt,psfig]{article}
\usepackage{amssymb}
\usepackage{amsmath,amsthm}
\usepackage{amsfonts}
\usepackage{graphicx}
\usepackage{graphics}
\numberwithin{equation}{section}

\begin{document}
\title{\setlength{\baselineskip}{0.6\baselineskip}
\begin{normalsize}\textbf{TOPOLOGICAL QUANTUM CODES FROM SELF-COMPLEMENTARY SELF-DUAL GRAPHS}\end{normalsize}}
\author{\footnotesize{AVAZ NAGHIPOUR}\\
\footnotesize{\it Department of Computer Engineering, University
College of Nabi Akram,}\\  \footnotesize{\it No. 1283 Rah Ahan
Street, Tabriz, Iran}\\
\footnotesize{\it Department of Applied Mathematics,
Faculty of Mathematical Sciences, University of Tabriz,}\\
\footnotesize{\it 29 Bahman Boulevard, Tabriz, Iran }\\
\footnotesize{\it a\_naghipour@tabrizu.ac.ir}\\
[2mm] \footnotesize{
 MOHAMMAD ALI JAFARIZADEH}\\
\footnotesize{\it Department of Theoretical Physics and
Astrophysics, Faculty of Physics, University of Tabriz,}\\
\footnotesize{\it 29 Bahman Boulevard, Tabriz, Iran }\\
\footnotesize{\it jafarizadeh@tabrizu.ac.ir}\\
[2mm] \footnotesize{
 SEDAGHAT SHAHMORAD}\\
\footnotesize{\it Department of Applied Mathematics,
Faculty of Mathematical Sciences, University of Tabriz,}\\
\footnotesize{\it 29 Bahman Boulevard, Tabriz, Iran }\\
\footnotesize{\it shahmorad@tabrizu.ac.ir}}
\date{\footnotesize{19 March 2015}}
\maketitle
\begin{small}
\hspace{-0.7cm}
\begin{quote}
\footnotesize{In this paper we present two new classes of binary
quantum codes with minimum distance of at least three, by
self-complementary self-dual orientable embeddings of ``voltage
graphs'' and ``Paley graphs in the Galois field $GF(p^{r})$'', where
$p\in \mathbb{P}$ and $r\in \mathbb{Z}^{+}$. The parameters of two
new classes of quantum codes are
$[[(2k'+2)(8k'+7),2(8k'^{2}+7k'),d_{min}]]$ and
$[[(2k'+2)(8k'+9),2(8k'^{2}+9k'+1),d_{min}]]$ respectively, where
$d_{min}\geq 3$. For these quantum codes, the code rate approaches
$1$ as $k'$ goes to infinity.}
\vspace{.3cm}
\end{quote}
\parindent 1em
\end{small}
\section{\hspace*{-0.6cm}{.} Introduction}
Quantum error-correcting codes (QEC) plays an important role in the
theory of quantum information and computation. A main difficult to
realize quantum computation is decoherence of quantum bits due to
the interaction between the system and the surrounding environments.
The QEC provide an efficient way to overcome decoherence. The first
quantum code $[[9,1,3]]$ was discovered by Shor [1]. Calderbank et
al. [2] have introduced a systematic way for constructing the QEC
from classical error-correcting code. The problem of constructing
toric quantum codes has motivated considerable interest in the
literature. This problem was generalized within the context of
surface codes [8] and color codes [3]. The most popular toric code
was the first proposed by Kitaev's [5]. This code defined on a
square lattic of size $m\times m$ on the torus. The parameters of
this class of codes are $[[n,k,d]]=[[2m^2,2,m]]$. In the similar
way, the authors in [7] have introduced a construction of
topological quantum codes in the projective plane $\mathbb{R}P^2$.
They showed that the original Shor's $9$-qubit repetition code is
one of these codes which can be constructed in a planar domain.
\\
\hspace*{0.5cm} Leslie in [6] proposed a new type of sparse CSS
quantum error correcting codes based on the homology of hypermaps
defined on an $m\times m$ square lattice. The parameters of
hypermap-homology codes are $[[(\frac{3}{2})m^2,2,m]]$. These codes
are more efficient than Kitaev's toric codes. This seemed to suggest
that good quantum codes maybe constructed by using hypergraphs. But
there are other surface codes with better parameters than the
$[[2m^2,2,m]]$ toric code. There exist surface codes with parameters
$[[m^2+1,2,m]]$, called homological quantum codes. These codes were
introduced by Bombin and Martin-Delgado [8].
\\
\hspace*{0.5cm} Authors in [9] presented a new class of toric
quantum codes with parameters $[[m^2,2,m]]$, where $m=2(l+1),l\geq
1$. Sarvepalli [10] studied relation between surface codes and
hypermap-homology quantum codes. He showed that a canonical hypermap
code is identical to a surface code while a noncanonical hypermap
code can be transformed to a surface code by CNOT gates alone. Li et
al. [17] were given a large number of good binary quantum codes of
minimum distances five and six by Steane's Construction. In [18]
good binary quantum stabilizer codes are obtained via graphs of
Abelian and non-Abelian groups schemes. In [19], Qian presented a
new method of constructing quantum codes from cyclic codes over
finite ring $\mathrm{F_{2}}+v\mathrm{F_{2}}$.
\\
\hspace*{0.5cm} Our aim in this work is to present two new classes
of binary quantum codes with parameters
$[[(2k'+2)(8k'+7),2(8k'^{2}+7k'),d_{min}]]$ and
$[[(2k'+2)(8k'+9),2(8k'^{2}+9k'+1),d_{min}]]$ respectively, based on
results of Hill in self-complementary self-dual graphs [13]. Binary
quantum codes are defined by pair $(H_{X},H_{Z})$ of
$\mathbb{Z}_{2}$-matrices with $H_{X}H_{Z}^{T}=0$. These codes have
parameters $[[n,k,d_{min}]]$, where $k$ logical qubits are encoded
into $n$ physical qubits with minimum distance $d_{min}$. A minimum
distance $d_{min}$ code can correct all errors up to
$\lfloor\frac{d_{min}-1}{2}\rfloor$ qubits. The code rate for these
quantum codes of length $n=(2k'+2)(8k'+7)$ and $n=(2k'+2)(8k'+9)$ is
determined by $\frac{k}{n}=\frac{2(8k'^{2}+7k')}{(2k'+2)(8k'+7)}$
and $\frac{k}{n}=\frac{2(8k'^{2}+9k'+1)}{(2k'+2)(8k'+9)}$, and this
rate approaches $1$ as $k^{'}$ goes to infinity.
\\
\hspace*{0.5cm} The paper is organized as follows. The definition
simplices, chain complexes and homology group are recalled in
Section 2. In Section 3 we shall briefly present the voltage graphs
and their derived graphs. In Section 4, we give a brief outline of
self-complementary self-dual graphs. Section 5 is devoted to present
new classes of binary quantum codes by using self-complementary
self-dual orientable embeddings of voltage graphs and Paley graphs.
The paper is ended with a brief conclusion.
\section{\hspace*{-0.6cm}{.} Homological algebra}
In this section, we review some fundamental notions of homology
spaces. For more detailed information about homology spaces, refer
to [4], [12].
\\
\\
\textbf{Simplices.\hspace*{1mm}} Let $m,n\in \mathbb{N}$, $m\geq n$.
Let moreover the set of points
$\{\upsilon_{0},\upsilon_{1},...,\upsilon_{n}\}$ of $\mathbb{R}^m$
be geometrically independent. A $n$-simplex $\Delta$ is a subset of
$\mathbb{R}^m$ given by
\\
\begin{equation}
\Delta=\{x\in\mathbb{R}^m|x=\sum_{i=0}^{n}t_{i}\upsilon_{i}; 0\leq
t_{i}\leq 1; \sum_{i=0}^{n}t_{i}=1\}.
\end{equation}
\\
\\
\textbf{Chain complexes.\hspace*{1mm}} Let $K$ be a simplicial
complex and $p$ a dimension. A $p$-\textit{chain} is a formal sum of
$p$-simplices in $K$. The standard notation for this is
$c=\sum_{i}n_{i}\sigma_{i}$, where $n_{i}\in \mathbb{Z}$ and
$\sigma_{i}$ a $p$-simplex in $K$. Let $C_{p}(K)$ be the set of all
$p$-chains on $K$. The \textit{boundary homomorphism} $\partial_{p}:
C_{p}(K)\longrightarrow C_{p-1}(K)$ is defined as
\\
\begin{equation}
\partial_{k}(\sigma)=\sum_{j=0}^{k}(-1)^j[\upsilon_{0},\upsilon_{1},...,\upsilon_{j-1},\upsilon_{j+1},...,\upsilon_{k}].
\end{equation}
\\
The \textit{chain complex} is the sequence of chain groups connected
by boundary homomorphisms,
\\
\begin{equation}
\cdots
\stackrel{\partial_{p+2}}{\longrightarrow}C_{p+1}\stackrel{\partial_{p+1}}{\longrightarrow}C_{p}
\stackrel{\partial_{p}}{\longrightarrow}C_{p-1}\stackrel{\partial_{p-1}}{\longrightarrow}\cdots
\end{equation}
\\
\textbf{Cycles and boundaries.\hspace*{1mm}} We are interested in
two subgroups of $C_{p}(K)$, \textit{cycle} and \textit{boundary}
groups. The $p$-th cycle group is the kernel of
$\partial_{p}:C_{p}(K)\longrightarrow C_{p-1}(K)$, and denoted as
$Z_{p}=Z_{p}(K)$. The $p$-th boundary group is the image of
$\partial_{p+1}:C_{p+1}(K)\longrightarrow C_{p}(K)$, and denoted as
$B_{p}=B_{p}(K)$.
\\
\\
\textbf{Definition 2.1\hspace*{1mm}}(Homology group, Betti number).
The $p$-\textit{th homology group} $H_{p}$ is the $p$-th cycle group
modulo the $p$-th boundary group, $H_{p}=Z_{p}/B_{p}$. The
$p$-\textit{th Betti number} is the rank (i.e. the number of
generators) of this group, $\beta_{p}$=rank $H_{p}$. So the first
homology group $H_{1}$ is given as
\\
\begin{equation}
H_{1}=Z_{1}/B_{1}.
\end{equation}
\\
From the algebraic topology, we can see that the group $H_{1}$ only
depends, up to isomorphisms, on the topology of the surface [4]. In
fact
\\
\begin{equation}
H_{1}\simeq \mathbb{Z}_{2}^{2g}.
\end{equation}
\\
where $g$ is the genus of the surface, i.e. the number of ``holes''
or ``handles''. We then have
\\
\begin{equation}
|H_{1}|=2^{2g}.
\end{equation}
\section{\hspace*{-0.6cm}{.} Voltage graphs and their derived graphs}
Let $G=(V,E)$ be a multigraph for which every edge has been assigned
a direction, and $\mathcal{V}$ be a finite group. A \textit{voltage
assignment} of $G$ in $\mathcal{V}$ is a function
$\alpha:E\rightarrow\mathcal{V}$, that labels the arcs of $G$ with
elements of $\mathcal{V}$. The triple $(G,\mathcal{V},\alpha)$ is
called an (ordinary) voltage graph. The \textit{derived graph}
(\textit{lift}, or \textit{covering}) $G'=(V',E')$ (also denoted
$G^{\alpha}$), is defined as follows:
\\
\\
i) $V'=V\times\mathcal{V}$
\\
\\
ii) If $e=(a,b)\in E$ where $a,b\in V$ and $\alpha(e)=v$ for some
$v\in \mathcal{V}$, then $(e,u)=e_u= \hspace*{4mm} (a_u,b_{uv})\in
E'$ where $a_u,b_{uv}\in V'$ for all $u\in\mathcal{V}$.
\\
\\
\textbf{Definition 3.1.\hspace*{1mm}}Let $\mathbb{Z}^{t}$ denote
$\underbrace{\mathbb{Z}_{2}\times \mathbb{Z}_{2}\times
\cdots\times\mathbb{Z}_{2}}_{t}$. A \textit{binary vector} has even
weight if it has an even number of $1$'s and has odd weight
otherwise. An $\varepsilon$-vector is a vector in
$\mathbb{Z}_{2}^{t-1}$ with even weight. A $\sigma$-vector is a
vector in $\mathbb{Z}_{2}^{t-1}$ with odd weight. We label the
$\varepsilon$-vectors so that
$\varepsilon_{1}<\varepsilon_{2}<\cdots<\varepsilon_{2^{t-2}}$.
Similarly, label the $\sigma$-vectors so that
$\sigma_{1}<\sigma_{2}<\cdots<\sigma_{2^{t-2}}$.
\\
\\
\textbf{Definition 3.2.\hspace*{1mm}} A $1\varepsilon$-vector is a
vector in $\mathbb{Z}_{2}^{t}$ where the first entry is a one and
the remainder of the vector is an $\varepsilon$-vector. The
$1\sigma$-, $0\varepsilon$- and $0\sigma$-vectors can be defined in
a similar fashion. A $1\varepsilon$-edge is an edge with a
$1\varepsilon$-vector as a voltage assignment. The $1\sigma$-,
$0\varepsilon$- and $0\sigma$-edges can be defined in a similar
fashion. For example, when $t=3$, we have the following table:
\\
$$
\begin{tabular}{|c|c|}
  \hline
  $0\varepsilon_{1}=000=0$ & $0\sigma_{1}=001=1$\\
  $0\varepsilon_{2}=011=3$ & $0\sigma_{2}=010=2$\\
  \hline
  $1\varepsilon_{1}=100=4$ & $1\sigma_{1}=101=5$\\
  $1\varepsilon_{2}=111=7$ & $1\sigma_{2}=110=6$\\
  \hline
\end{tabular}
$$
\\
\textbf{Definition 3.3.\hspace*{1mm}} A \textit{link} is an edge
which is incident with $2$ different vertices. A \textit{loop} is an
edge which has two incidences with the same vertex. A \textit{half
edge} is an edge together with one of its incident vertices.
\\
\\
\textbf{Definition 3.4.\hspace*{1mm}} Let $t\geq 3$. Let $H_{t}$ be
a voltage graph defined as follows over the group
$(\mathbb{Z}_{2}^{t},\oplus)$; $H_{t}$ has two vertices, $u$ and
$v$. There are $2^{t-1}$ links between $u$ and $v$, with voltage
assignments $0\sigma_{1},\ldots,0\sigma_{2^{t-2}}$ and
$1\varepsilon_{1},\ldots,1\varepsilon_{2^{t-2}}$ (equivalently, all
possible vectors in $\mathbb{Z}_{2}^{t}$ with odd weight). There are
$2^{t-1}$ half edges about $v$ with voltage assignments
$0\sigma_{1},\ldots,0\sigma_{2^{t-2}}$ and
$1\sigma_{1},\ldots,1\sigma_{2^{t-2}}$. Similarly, there are
$2^{t-1}-1$ half edges about $u$ with voltage assignments
$0\varepsilon_{2},0\varepsilon_{3},\ldots,0\varepsilon_{2^{t-2}}$
and $1\varepsilon_{1},\ldots,1\varepsilon_{2^{t-2}}$.
\section{\hspace*{-0.6cm}{.} Self-complementary self-dual graphs}
Let $G=(V,E)$ be a simple graph. The complement $\overline{G}$ of G
has the same vertices as $G$, and every pair of vertices are
adjacent by an edge in $\overline{G}$ if and only if they are not
adjacent in $G$. A graph $G$ is self-complementary if
$G\cong\overline{G}$. Let $M=(V,E,F)$ be a fixed map of $G$, with
dual map $M^{*}=(F^{*},E^{*},V^{*})$. $M$ is graphically self-dual
if $(V,E)\cong(F^{*},E^{*})$.
\\
\\
\textbf{Theorem 4.1.\hspace*{1mm}} If $G$ is a self-complementary
graph on $m$ vertices, then $|E(G)|=\frac{m(m-1)}{4}$, and
$m\equiv0$ or $1$ (mod $4$).
\\
\\
\textbf{Proof.\hspace*{1mm}} See [14].
\\
\\
\textbf{Theorem 4.2.\hspace*{1mm}} If $G$ is a self-complementary
self-dual graph on $m$ vertices with a self-dual embedding on an
orientable surface of genus $g$, then $m\equiv0$ or $1$ (mod $8$).
In particular, if $m=8+8k'$, then $g=8(k'^{2})+7k'$, and if
$m=9+8k'$, then $g=8(k'^{2})+9k'+1$.
\\
\\
\textbf{Proof.\hspace*{1mm}} See [13].
\section{\hspace*{-0.6cm}{.} Quantum codes from graphs on surfaces}
The idea of constructing CSS (Calderbank-Shor-Steane) codes from
graphs embedded on surfaces has been discussed in a number of
papers. See for detailed descriptions e.g. [11]. Let $X$ be a
compact, connected, oriented surface (i.e. 2-manifold) with genus
$g$. A tiling of $X$ is defined to be a cellular embedding of an
undirected (simple) graph $G=(V,E)$ in a surface. This embedding
defines a set of faces $F$. Each face is described by the set of
edges on its boundary. This tiling of surface is denoted
$M=(V,E,F)$. The dual graph $G$ is the graph $G^*=(V^*,E^*)$ such
that:
\\
\\
i) One vertex of $G^*$ inside each face of $G$,
\\
\\
ii) For each edge $e$ of $G$ there is an edge $e^*$ of $G^*$ between
the two vertices of \hspace*{4mm} $G^*$ corresponding to the two
faces of $G$ adjacent to $e$.
\\
\\
It can be easily seen that, there is a bijection between the edges
of $G$ and the edges of $G^*$.
\\
\\
\hspace*{0.5cm} There is an interesting relationship between the
number of elements of a lattice embedded in a surface and its genus.
The Euler characteristic of $X$ is defined as its number of vertices
($|V|$) minus its number of edges ($|E|$) plus its number of faces
($|F|$), i.e.,
\\
\begin{equation}
\chi=|V|-|E|+|F|.
\end{equation}
\\
For closed orientable surfaces we have
\\
\begin{equation}
\chi=2(1-g).
\end{equation}
\\
The \textit{surface code} associated with a tiling $M=(V,E,F)$ is
the CSS code defined by the matrices $H_{X}$ and $H_{Z}$ such that
$H_{X}\in \mathcal{M}_{|V|,|E|}(\mathbb{Z}_{2})$ is the vertex-edge
incidence matrix of the tiling and $H_{Z}\in
\mathcal{M}_{|F|,|E|}(\mathbb{Z}_{2})$ is the face-edge incidence
matrix of the tiling. Therefore, from $(X,G)$ is constructed a CSS
code with parameters $[[n,k,d]]$. where $n$ is the number of edges
of $G$, $k=2g$ (by (2.6)) and $d$ is the shortest non-boundary cycle
in $G$ or $G^*$. In this work, the minimum distance of quantum codes
by a parity check matrix $H$ (or generator matrix) is obtained. For
a detailed information to compute the minimum distance, we refer the
reader to [15].
\subsection{\hspace*{-0.5cm}{.} New class of $[[(2k'+2)(8k'+7),2(8k'^{2}+7k'),d_{min}]]$ binary quantum codes from
embeddings of voltage graphs} Our aim in this subsection is to
construct new class of binary quantum codes by using
self-complementary self-dual orientable embeddings of voltage
graphs. Let $G_{t}$ be the lift of voltage graph $H_{t}$ defined
over the group $(\mathbb{Z}_{2}^{t},\oplus)$. Since
$|V(G_{t})|=|V(H_{t})|\times|\mathbb{Z}_{2}^{t}|=2\times2^{t}$, for
$t=3$, $m=|V(G_{t})|=2^{3}\times2=2^{4}\equiv0$ (mod 8). On the
other hand, since by Theorems in Section 4,
$|E(G)|=\frac{m(m-1)}{4}$ and $m=8+8\times1$, thus $|E(G)|=60$ and
$g=15$. From Definition 3.4 we get the following adjacency matrix
for $t=3$:
\\
$$
 A= \left(
     \begin{array}{lc}
       IXX+XII+XXX & XII+IXI+IIX+XXX \\ \\
       XII+IXI+IIX+XXX & IIX+IXI+XIX+XXI \\
      \end{array}
   \right)
$$
\\
where $I$ is an $2\times 2$ identity matrix and $X$ is an Pauli
matrix. Also, we will sometimes use notation where we omit the
tensor signs. For example $IXX$ is shorthand for $I\otimes X\otimes
X$. After finding the vertex-edge incidence matrix $H_{X}$ using the
above adjacency matrix and the face-edge incidence matrix $H_{Z}$ by
Gaussian elimination and the \textit{standard form} of the parity
check matrix in [15], one can be easily seen that $H_{X}H_{Z}^{T}=0$
and $d_{min}=3$. Therefore, the code with parameters $[[60,30,3]]$
is constructed.
\\
\\
In general, the adjacency matrix $A=(a_{ij})_{2^{t+1}\times
2^{t+1}}$ of derived voltage graph by Definition 3.4, is
\\
$$
 A= \left(
     \begin{array}{cc}
      B & C \\ \\
      C & D \\
      \end{array}
   \right)
$$
\\
where
\\
\begin{equation*}
B= \frac{1}{2}(I+X)\otimes\{\underbrace{(I+X)\otimes(I+X)\otimes
\cdots\otimes(I+X)}_{t-1}+
\end{equation*}
\begin{equation*}
+\underbrace{(I-X)\otimes(I-X)\otimes
\cdots\otimes(I-X)}_{t-1}\}-\underbrace{I\otimes I\otimes
\cdots\otimes I}_{t};
\end{equation*}
\\
\begin{equation*}
\hspace*{-14mm} C= \frac{1}{2}\{\underbrace{(I+X)\otimes(I+X)\otimes
\cdots\otimes(I+X)}_{t}-
\end{equation*}
\begin{equation*}
\hspace*{-28mm} -\underbrace{(I-X)\otimes(I-X)\otimes
\cdots\otimes(I-X)}_{t}\};
\end{equation*}
\\
\begin{equation*}
D= \frac{1}{2}(I+X)\otimes\{\underbrace{(I+X)\otimes(I+X)\otimes
\cdots\otimes(I+X)}_{t-1}-
\end{equation*}
\begin{equation*}
\hspace*{-27mm} -\underbrace{(I-X)\otimes(I-X)\otimes
\cdots\otimes(I-X)}_{t-1}\}.
\end{equation*}
\\
\\
With finding the matrix $H_{X}$ using the above adjacency matrix
$A=(a_{ij})_{2^{t+1}\times 2^{t+1}}$ and the matrix $H_{Z}$ by
Gaussian elimination and the standard form of the parity check
matrix, the code minimum distance of at least three is obtained.
\\
\\
After determining $d_{min}$, by using the Theorems in Section $4$
the class of codes with parameters
$[[(2k'+2)(8k'+7),2(8k'^{2}+7k'),d_{min}]]$, $k'\geq 1$ is
constructed.
\subsection{\hspace*{-0.5cm}{.} New class of $[[(2k'+2)(8k'+9),2(8k'^{2}+9k'+1),d_{min}]]$ binary \hspace*{-3mm} quantum codes from
embeddings of Paley graphs} The construction of this class will be
based on self-complementary self-dual orientable embeddings of Paley
graphs in the Galois field $GF(p^{r})$, where $p\in \mathbb{P}$ and
$r\in \mathbb{Z}^{+}$.
\\
\\
\textbf{Definition 5.2.1.\hspace*{1mm}} Let $G$ be a group and $S$
be a subset of $G\backslash\{id\}$. We say that a graph $X$ is a
\textit{Cayley graph} with \textit{connection set} $S$, written
$X$=Cay$(G,S)$, if
\\
\\
i) $V(X)=G$,
\\
\\
ii) $E(X)=\{\{g,sg\}|g\in G, s\in S\}$.
\\
\\
\textbf{Definition 5.2.2.\hspace*{1mm}} Let $m=p^{r}\equiv1$ (mod
$8$), $p\in \mathbb{P}$ and $r\in \mathbb{Z}^{+}$. A \textit{Paley
graph} is a cayley graph $P_{m}$=Cay$(X_{m},\Delta_{m})$, where
$X_{m}=\underbrace{\mathbb{Z}_{p}\times \mathbb{Z}_{p}\times
\cdots\times\mathbb{Z}_{p}}_{m}$ is the additive group of the Galois
field $GF(p^{r})$ and $\Delta_{m}=\{1,x^{2},x^{4},\ldots, x^{m-3}\}$
for a primitive element $x$ of $GF(p^{r})$.
\\
\\
Let $G=(V,E)$ be a self-complementary self-dual graph on $m$
vertices. From Theorem 4.1, we know that $|E(G)|=\frac{m(m-1)}{4}$.
Also, from Theorem 4.2 and Definition 5.2.2, with a self-dual
embedding on an orientable surface of genus $g$, we know that if
$m=9+8k'\equiv1$ (mod $8$), then $g=8(k'^{2})+9k'+1$. Therefore,
$|E(G)|=\frac{(9+8k')(8+8k')}{4}=(9+8k')(2+2k')$. Since in this
self-dual embedding on an orientable surface the code minimum
distance is at least three. Thus the code parameters are given by:
the code minimum distance is $d_{min}\geq 3$; the code length is
$n=|E(G)|=(9+8k')(2+2k')$ and $k=2g=2(8k'^{2}+9k'+1)$. Consequently,
the class of codes with parameters
$[[(2k'+2)(8k'+9),2(8k'^{2}+9k'+1),d_{min}]]$, $k'\geq 0$ is
obtained.
\\
\\
\textbf{Example 5.2.1.\hspace*{1mm}} Let $m=3^{2}\equiv1$ (mod $8$).
Then $P_{9}$=Cay$(X_{9},\Delta_{9})$, where
$X_{9}=\mathbb{Z}_{3}\times \mathbb{Z}_{3}$ is the additive group of
the Galois field $GF(3^{2})$ and
$\Delta_{9}=\{1,x^{2},x^{4},x^{6}\}$ for a primitive element $x$ of
$GF(3^{2})$. In fact, $\Delta_{9}$ is the set of all squares in
$GF(3^{2})$. Let $p(x)\in \mathbb{Z}_{3}[x]$ be an irreducible
polynomial of degree $2$. Then the elements of
$\mathbb{Z}_{3}[x]/\langle p(x)\rangle$ will be polynomials of
degree $1$ or less and there will be $3^{2}=9$ such polynomials. So,
in terms of representatives, the elements of $GF(9)$ are
$\{ax+b|a,b\in \mathbb{Z}_{3}\}$. We denote these as:
\\
\begin{equation*}
g_{0}=0x+0 \hspace*{4mm} g_{3}=1x+0 \hspace*{4mm} g_{6}=2x+0
\end{equation*}
\begin{equation*}
g_{1}=0x+1 \hspace*{4mm} g_{4}=1x+1 \hspace*{4mm} g_{7}=2x+1
\end{equation*}
\begin{equation*}
g_{2}=0x+2 \hspace*{4mm} g_{5}=1x+2 \hspace*{4mm} g_{8}=2x+2
\end{equation*}
\\
Based on results of Conrad in finite fields [16], the monic
irreducible quadratics in $\mathbb{Z}_{3}[x]$ are $x^{2}+1$,
$x^{2}+x+2$ and $x^{2}+2x+2$. Let $p(x)=x^{2}+x+2$. Then $g_{3}=x$
is a generator of the nonzero elements in the field
$\mathbb{Z}_{3}[x]/\langle x^{2}+x+2\rangle$.
\\
\\
$g_{3}=x=g_{3}$\\ \\
$g_{3}^{2}=x^{2}=-x-2=2x+1=g_{7}$\\ \\
$g_{3}^{3}=x(2x+1)=2x^{2}+x=2(-x-2)+x=-x-1=2x+2=g_{8}$\\ \\
$g_{3}^{4}=x(2x+2)=2x^{2}+2x=2(-x-2)+2x=-4=2=g_{2}$\\ \\
$g_{3}^{5}=x(2)=2x=g_{6}$\\ \\
$g_{3}^{6}=x(2x)=2x^{2}=2(-x-2)=-2x-4=x+2=g_{5}$\\ \\
$g_{3}^{7}=x(x+2)=x^{2}+2x=-x-2+2x=x-2=x+1=g_{4}$\\ \\
$g_{3}^{8}=x(x+1)=x^{2}+x=-x-2+x=-2=1=g_{1}$
\\
\\
By Definitions in Subsection 5.2, we get the following adjacency
matrix for $GF(9)$:
\\
\\
\begin{equation*}
A=\left(
\begin{array}{ccccccccc}
0 & 1 & 1 & 0 & 0 & 1 & 0 & 1 & 0\\
1 & 0 & 1 & 1 & 0 & 0 & 0 & 0 & 1\\
1 & 1 & 0 & 0 & 1 & 0 & 1 & 0 & 0\\
0 & 1 & 0 & 0 & 1 & 1 & 0 & 0 & 1\\
0 & 0 & 1 & 1 & 0 & 1 & 1 & 0 & 0\\
1 & 0 & 0 & 1 & 1 & 0 & 0 & 1 & 0\\
0 & 0 & 1 & 0 & 1 & 0 & 0 & 1 & 1\\
1 & 0 & 0 & 0 & 0 & 1 & 1 & 0 & 1\\
0 & 1 & 0 & 1 & 0 & 0 & 1 & 1 & 0
\end{array}\right)
\end{equation*}
\\
\\
After finding the matrices $H_{X}$ and $H_{Z}$ using the Theorems in
Section 4, the code with parameters $[[18,2,3]]$ is obtained. Note
that the matrix $H_{Z}$ is given by Gaussian elimination and the
standard form of the parity check matrix in [15].
\\
\\
\section{\hspace*{-0.6cm}{.}\ Conclusion}
We have considered the presentation of two new classes of binary
quantum codes by using self-complementary self-dual orientable
embeddings of voltage graphs and Paley graphs. These codes is
superior to quantum codes presented in other references. We point
out the classes $[[(2k'+2)(8k'+7),2(8k'^{2}+7k'),d_{min}(\geq 3)]]$
and $[[(2k'+2)(8k'+9),2(8k'^{2}+9k'+1),d_{min}(\geq 3)]]$ of quantum
codes achieving the best ratio $\frac{k}{n}$.

\end{document}